\documentclass[12pt]{article}

\usepackage{graphics}

\input epsf
\newcommand{\be}{\begin{equation}}
\newcommand{\ee}{\end{equation}}

\newcommand{\ra}{\rightarrow}
\newcommand{\bea}{\begin{eqnarray}}
\newcommand{\eea}{\end{eqnarray}}

\newcommand{\p}{\partial}

\begin{document}

\begin{flushright}
UCLA/04/TEP/07\\
hep-th/0404073
\end{flushright}

\bigskip \bigskip
\centerline{\large \bf 
Splitting hairs of the three charge black hole.}
\bigskip
\bigskip
\centerline{{\bf Iosif Bena }}
\medskip
\centerline{ Department of Physics and Astronomy}
\centerline{University of California}
\centerline{Los Angeles, CA  90095, USA }
\bigskip
\centerline{{\rm iosif@physics.ucla.edu} }
\bigskip \bigskip

\begin{abstract}

We construct the large radius limit of the metric of three charge supertubes 
and three charge BPS black rings by using the fact that supertubes preserve the same 
supersymmetries as their component branes. Our solutions reproduce a few of the properties 
of three charge supertubes found recently using the Born Infeld description.
Moreover, we find that these solutions pass a number of rather nontrivial tests which 
they should pass if they are to describe some of the hair of three charge black 
holes and three charge black rings.  


\end{abstract}
\newpage

\section{Introduction}

Recently, there has been a renewed interest in the properties of 
supertubes \cite{supertube}, due mainly to their possible use in counting black hole
entropy. In the case of the D1-D5 system this has been successfully carried out  
\cite{mathurlunin,LuninJY,mathurstretch,LuninBJ,LuninIZ,mathur,MathurSV} by 
relating the various configurations of 
two charge supertubes to the supersymmetric ground states of the D1-D5 system.  
Moreover, the supergravity solutions of these supertubes have been found 
to be regular. This gives a one to one map between microstates of the D1-D5 system and 
classical supergravity solutions, which can thus be thought of as being 
the hair of this system. 

It is a rather fascinating possibility that this story extends to 
three charge black hole with a macroscopic horizon area. So far this has been 
directly checked only for a 
single unit of momentum \cite{mathur}.  Moreover, in \cite{bk} the three charge 
supertubes which might comprise the hair of the three charge black hole have 
been constructed\footnote{Other three charge supertubes have also appeared
in \cite{aspects}.} using both the Born-Infeld action of the dipole branes (generalizing the
construction in \cite{supertube,MateosPR}), and 
the nonabelian Born Infeld action of one of the constituent branes (generalizing 
the construction in \cite{f1,unp,bak}) . 
The properties of these three charge supertubes were found to be consistent 
with them describing some of the hair of the three charge black hole.
Moreover, it was argued that three charge supertubes with angular momentum higher 
than that of the BMPV black hole should represent the hair of a yet to be found BPS three 
charge black ring.

It is important therefore to find the supergravity solutions for
arbitrary three charge supertubes, and to see if these can be put
in one-to-one correspondence with the states of the D1-D5-P system. 
So far these solutions have been hard to find using standard solution generating 
techniques; it seems like more powerful methods are needed. 

Although in this paper we do not construct the most generic three charge supertube or 
BPS black ring solution, we construct the large radius (near tube) limit of these metrics, 
or otherwise put, the most generic metric sourced by a flat infinitely 
extended supertube. We hope in the near future to use 
these methods to find the full supertube and BPS black ring solutions. 

The three charge supertubes constructed in \cite{bk} have F1, D0 and D4 charges, as well 
as D2, D6 and NS5 dipole moments. Their Born Infeld description can only capture the D2 and 
D6 dipole moments, but not the NS5 one. Thus, if wants to count 
three charge supertubes (in the way  two charge supertubes have been counted \cite{bd}) 
it might not be enough to use the Born-Infeld action; one may have to count the BPS
 perturbations on the full near-tube geometry. 
Thus, finding this near-tube geometry is an important step towards the counting of the
most generic three charge supertubes. 

The crucial ingredient in our search for the near-supertube metric is the fact that the
supersymmetries preserved by the three charge supertube background and those of 
the three charge black hole are the same. This allows one to use a few of the rather 
powerful methods to find supersymmetric 
solutions that have been developed over the past few years 
by Warner, Nemeschansky, Pilch and collaborators \cite{nick}. Rather than solving 
complicated the 2'nd order equations, supersymmetric solutions
can be found by using the Killing spinors and their projectors to 
find algebraic first order relations between the 
metric and the forms, which are much easier to solve.

As we will see, tubes with three charges and three dipole charges come in two types. If the  
densities satisfy a certain relation, the geometry is smooth and has a zero size horizon.
As we will argue in section 4, this solution is the near tube limit of the 
supertube metric. If the charge densities are higher,  the metric has a horizon of nonzero 
area. This is nothing but the metric of the large 
radius limit of a BPS black ring. The existence of this black ring was 
conjectured \cite{bk} to account for the multiplicity of 
tubes with angular momentum larger  than that of a BMPV black hole; this is a first confirmation
of its existence.

Near a supertube with three 
charges and three dipole charges the metric is of the form $AdS_3 \times S^2 \times T^6$, 
in the duality frame in which the three charges are perpendicular M2 branes. The 
near horizon metric of the BPS black ring is of the form $AdS_2 \times S^2 \times S^1 
\times T^6$. Moreover, the Ricci scalar and other invariants are finite. 
Hence, both these geometries are smooth at $r=0$, and can be 
continued beyond this point. The fact that the geometry near the most 
generic three charge supertube is smooth 
is an important test which these supertubes must pass if they are to be 
the hair of the three charge black hole. We find that they indeed pass this test. 

We also show that the horizon area of the black ring is zero when
one supertube geometry 
exists, and starts increasing as the number of supertube configurations increases. 
The fact that a black ring with a nonzero horizon area appears exactly when the 
multiplicity of three charge supertube configuration becomes nonzero is a rather powerful 
confirmation of the picture of black hole microstates put forth in \cite{mathur,MathurSV}

In section 2 we find the flat supertube/black tube solution using the methods 
discussed above. In section 3 we discuss some of the special types of three charge supertubes,
including the tube with three charges and two dipole charges discussed in \cite{bk}. In 
section 4 we discuss the properties of the most generic supertubes (with three charges 
and three dipole charge), as well as the properties of the near horizon limit of the 
BPS black ring. We also discuss how the multiplicity of supertube configurations increases 
with the horizon area of the corresponding black tube, and explore the implications of this 
for black hole physics. We conclude in section 5. 

While this work was being prepared for publication, we became aware of \cite{lun}, 
where one of the metrics of the supertubes with three equal charges was found using the 
G-structure techniques of \cite{6d,g} in 6 dimensional minimal supergravity. It would be 
interesting to see if those techniques can be used to find the metrics of the 
most generic three charge supertubes and black tubes whose near horizon limit we give here.

\section{The large radius limit of the three charge supertube solution.}

The three charge supertubes which should be some of the hair of the D1-D5-P black hole can 
be related by U-duality to the D0-D4-F1 supertubes found in \cite{bk}, or to other 
configurations. In order to find the supergravity solution, it is useful to use a U-duality 
frame in which the symmetry between the three charges and the three dipole charges is 
manifest. Such a duality frame can be obtained by T-dualizing the configuration in \cite{bk} twice
along two of the $T^4$ directions, and lifting to 11 dimensions. This gives three M2 brane
charges, 
in the $123$, $145$, and $167$ planes, as well as three M5 brane dipole moments, in the $12345
\vartheta ,14567 \vartheta$, and $12367\vartheta$, directions. This duality frame has also the 
advantage of being free from the usual pathologies associated with boosting Kaluza Klein
monopoles. 

In the large radius/near supertube limit, the $\vartheta$ direction becomes infinitely 
extended -- we henceforth call this direction $z$. The three M5 brane dipole moments 
become M5 brane charges, and the angular momentum in the $\vartheta$ direction becomes 
momentum in the $z$ direction. The polar coordinates on the three directions transverse 
to the tube are  $r,\theta$ and $\phi$.

A generic two charge supertube can have a shape given by an arbitrary closed 
curve in the transverse space, and can also have arbitrary brane densities along 
this curve, as long as certain constraints are observed. The solutions we construct 
are for this curve being an infinite straight line, and for arbitrary local 
densities of the three charges. Thus, we allow the metric and forms to vary both 
along the tube (in the $z$ direction) and radially, but we require them to be SO(3) invariant.

We take the metric to have Lorentz frames of the form:
\bea
e^1 &=& e^{-2{A_1}(r,z) - 2{A_2}(r,z) - 2{A_3}(r,z)}
   \left( {dx}_{1} + k(r,z) {dz} \right), \\
e^2 &=&  e^{-2{A_1}(r,z) + {A_2}(r,z) +   {A_3}(r,z)} {dx}_{2},\\
e^3 &=&  e^{-2{A_1}(r,z) + {A_2}(r,z) +   {A_3}(r,z)} {dx}_{3},\\
e^4  &=& e^{{A_1}(r,z) - 2{A_2}(r,z) + {A_3}(r,z)}{dx}_{4}, \\
e^5  &=& e^{{A_1}(r,z) - 2{A_2}(r,z) + {A_3}(r,z)}{dx}_{5}, \\
e^6 &=& e^{{A_1}(r,z) + {A_2}(r,z) - 2{A_3}(r,z)}{dx}_{6}, \\
e^7 &=& e^{{A_1}(r,z) + {A_2}(r,z) - 2{A_3}(r,z)}{dx}_{7}, \\
e^8 &=&  e^{{A_1}(r,z) + {A_2}(r,z) +{A_3}(r,z)}{dz},\\
 e^9 &=&   e^{{A_1}(r,z) + {A_2}(r,z) + {A_3}(r,z)}\, {dr} , \\
e^{10} &=&   e^{{A_1}(r,z) + {A_2}(r,z) + {A_3}(r,z)} \, r  {d\theta } \\
e^{11} &=&    e^{{A_1}(r,z) + {A_2}(r,z) +  {A_3}(r,z)}\,  r \sin \theta  {d\phi }
\eea

The three M2 brane charges are in the 23,45 and 67 planes. In the flat tube limit of the metric,
everything is a function of $r$ and $z$ only. The harmonic functions $Z_i$ which are commonly  
used to write this metric are
\be
Z_i = e^{6 A_i}
\ee

Besides the electric 4-forms which we expect for smeared M2 branes, we also have a few
magnetic 
4-forms which correspond to the M5 charges of the supertube:
\bea
F_{123r} &= & -6{A_1}^{(1,0)}{e^{-6{A_1}}} \label{f1}\\
F_{145r} &= & -6{A_2}^{(1,0)}{e^{-6{A_2}}} \\
F_{167r} &= & -6{A_3}^{(1,0)}{e^{-6{A_3}}} \\
F_{123z} &= & -6{A_1}^{(0,1)}{e^{-6{A_1}}} \\
F_{145z} &= & -6{A_2}^{(0,1)}{e^{-6{A_2}}} \\
F_{167z} &= & -6{A_3}^{(0,1)}{e^{-6{A_3}}} \\
F_{23zr} &= & 2 t_1 \\
F_{45zr} &= & 2 t_3 \\
F_{67zr} &= & 2 t_3 \label{f9}\\
F_{23~10~11} &=& 2 d_1 \sin \theta     \\
F_{45~10~11} &=& 2 d_2 \sin \theta \\
F_{67~10~11} &=& 2 d_3 \sin \theta 
\label{f12}
\eea
In the case of the 2-charge supertube, $A_3 = d_1 =d_2 = t_i =0$. For no supertubes, 
$ d_i  = t_i  = k = 0$. Since everything depends on $r$ and $z$, the Bianchi identities for the
4-forms in equations (\ref{f1}-\ref{f9}) are automatically satisfied.  For $F_{ij~10~11}$ these
identities imply that  $d_1,d_2$ and $d_3$ are constant.

For no supertube, each of the three M2 branes present breaks some of the supersymmetry. The 4
remaining Killing spinors are
\be
\epsilon_i = e^{-A_1 - {A_2} -  {A_3}} \eta_i
\ee
where $\eta_i$ is a constant spinor annihilated by the three M2 brane projectors
\be
(1+ \Gamma^{123}) \eta_i = (1+ \Gamma^{145}) \eta_i =(1+ \Gamma^{167}) \eta_i =0
\ee

The crucial ingredient used in constructing this solution is the fact that 
the supertube preserves the same supersymmetry as its components. 
Therefore  the above Killing spinors should also be Killing spinors of the supertube 
background.  It is a straightforward exercise to use the supersymmetry 
variations, and find that in order for this to happen, the magnetic field 
strengths and $k$ should be related by:
\bea
-r^2 t_1 e^{6 A_1}    &=&     { d_2   e^{6 A_2} +d_3 e^{6A_3} + 3r^2 k A_1^{(1,0)}}, \label{k0} \\
- r^2 t_2 e^{6 A_2}   & =&     { d_1   e^{6 A_1} +d_3 e^{6A_3} + 3r^2 k A_2^{(1,0)}},  \\
-r^2 t_3 e^{6 A_3}    &=  &   { d_2   e^{6 A_2} +d_1 e^{6A_1} + 3r^2 k A_3^{(1,0)}},  \\
- r^2 k^{(1,0)} &= &  {2 \left(d_1 e^{6 A_1} + d_2 e^{6 A_2} + d_3 e^{6 A_3} \right) }
\label{k}
\eea

In principle the Bianchi  identities and the supersymmetry transformations 
are enough to determine that the background is a solution. However, in the case 
of the usual D-branes these equations give a solution sourced by an arbitrary 
distribution of branes. To find the solution for empty space with at most point sources one must 
use some of the equations of motion as well.

The easiest equations of motion to use come from $d * F = F \wedge F$, and give three other
equations:
\bea
A_1^{(0,2)} + A_1^{(2,0)}  &=& \frac{4 d_2 d_3 e^{-6 A_1} -  6 r^3 A_1^{(1,0)} 
- 18 r^4  \left(A_1^{(1,0)} \right)^2}{3 r^4} \\
A_2^{(0,2)} + A_2^{(2,0)}   &=& \frac{4 d_1 d_3 e^{-6 A_2} -  6 r^3 A_2^{(1,0)}  
- 18 r^4  \left(A_2^{(1,0)} \right)^2}{3 r^4} \\
A_3^{(0,2)} + A_3^{(2,0)}   &=& \frac{4 d_2 d_3 e^{-6 A_3} -  6 r^3 A_3^{(1,0)}  
- 18 r^4  \left(A_3^{(1,0)} \right)^2}{3 r^4},
 \eea
which imply that the functions $Z_i \equiv e^{6 A_i}$ obey the equations:
\bea
{\p^2 Z_1 \over \p z^2} + {\p^2 Z_1 \over \p r^2} +   
{2 \over r} {\p Z_1 \over \p r} &=& {8 d_2 d_3 \over r^4} \label{har1}\\ 
{\p^2 Z_2 \over \p z^2} +{\p^2 Z_2 \over \p r^2} +   
{2 \over r} {\p Z_2 \over \p r}  &=& {8 d_1 d_3 \over r^4} \\ 
{\p^2 Z_3 \over \p z^2} +{\p^2 Z_3 \over \p r^2} +  
 {2 \over r} {\p Z_3 \over \p r} &=& {8 d_2 d_1 \over r^4} \label{har3} .
\eea
The left hand side of this equation is nothing but the  harmonic equation in four dimensions. 
For no supertubes, the right hand side of each of the three equations above is zero, and the
metric is the one sourced by an arbitrary smeared distribution of the 
three types of M2 branes present.  

In the case of a two charge supertube, two of the dipole moments are zero, so the 
functions $Z_i$ are again harmonic. The generic solution \cite{emparan,mathurlunin,LuninIZ} is
simply 
given by smearing these harmonic functions along the curves which determine the shape of the
tube.

When two of the dipole charges are nonzero, the functions $Z_i$ are harmonic away 
from the tube; however near the tube the right hand side of (\ref{har1} - \ref{har3}) 
becomes important. Therefore, this solution, as well as the other three charge supertube 
solutions cannot be obtained from harmonic brane solutions by usual solution generating 
techniques.

If the charge densities are independent on the $z$ direction, the equations above have very simple
asymptotically flat solutions:

\bea
Z_1 &=& \frac{4 d_2 d_3}{r^2} +  \frac{ Q_1}{r}   + 1 \\
Z_2 &=&\frac{4 d_1 d_3}{r^2} +  \frac{ Q_2}{r}  +  1\\
Z_3 &=& \frac{4 d_2 d_1}{r^2} +  \frac{ Q_3}{r}  + 1 \\
{\p k \over \p r} &= & -  {2 \over r^2}  \left(d_1 Z_1 + d_2 {Z_2} + d_3 {Z_3} \right) \\
k &=& +{8 d_1 d_2 d_3 \over r^3} + {d_1 Q_1 + d_2 Q_2 + d_3 Q_3 \over r^2} + 
{2 d_1+2 d_2+ 2 d_3 \over r}
\label{k2}
\eea

When the charge densities have a nontrivial dependence along the tube directions, 
the solutions are more involved, and possibly quite interesting --  among them might
be for example a solution with several 3 charge black holes connected by a straight tube. 
We hope to explore those and other solutions in the future.

\section{The physics of the solutions - special cases}

As we have seen, when only one dipole charge is present, the functions $Z_i$ are harmonic, 
which explains why the two charge supertube solutions found in 
\cite{emparan,mathurlunin,LuninIZ} are 
given by smearing the source of the harmonic function over the curve which gives the tube shape.

When two or three of the dipole charges of the supertube are nonzero, 
the rotation parameter $k$ given by (\ref{k}) is always such that $g_{zz}$ is zero near 
the branes. This also happens for the two charge supertubes, and should be expected,
 since the $z$ direction is common to the 3 M5 branes in the problem. .  

Since the $z$ direction is the direction along the tube, and is periodically identified, 
if $g_{zz} $ becomes negative inside a region, then the solutions have 
closed timelike curves. The absence of these curves requires: 
\be
k^2 \leq Z_1 Z_2 Z_3.
\label{ctc}
\ee
Like in the two charge supertube case, this relation determines a lower  bound on the densities of
the three charges.

In the following subsections we will be focusing on tubes where the local densities do not 
depend on $z$; hence the solution is given by (\ref{k2}). 

\subsection{ The two charge supertube -- two charges and one dipole charge}

In this case 
\be
Z_3=1, ~~~ Z_1 = 1+ {Q_1 \over r }, ~~~Z_2 = 1+ {Q_2 \over r }, ~~~ k = {2 d_3 \over r}
\ee
This reproduces the near tube form of the metric found in \cite{emparan,LuninIZ}. The condition
that the background be free from closed timelike curves implies $4 d_3 ^2 \leq Q_1 Q_2$.
However, as we know from the D-brane physics of the tube, and as also found in the analysis of
\cite{LuninIZ}, the inequality must be saturated in order for the solution to be regular. Thus, the
two charge tubes satisfy
\be
4 d_1 ^2 = Q_2 Q_3.
\label{2c}
\ee
If the charges are higher, the solution is that of coincident supertubes and regular branes, 
and is probably not a bound state.

\subsection{ Three charges and two dipole charges}

The tubes with three charges and two dipole charges have been explored in 
\cite{bk} using the Born Infeld action. The tubes with a minimum number of dipole branes (which
are the largest, and can be thought of as the building blocks of tubes with higher dipole charges) 
were found to have the property
\be
{Q_2^D \over Q_0} = {Q_6^D \over Q_4}
\label{ratio}
\ee
where $Q_0$ and $Q_4$ are the local zero brane and 4-brane charge densities. It is quite 
remarkable that the same relation comes up if we require the near tube geometry to be 
free of closed timelike curves.
Indeed, near the tube the leading terms in the harmonic functions are 
\be
Z_1 \approx {4 d_2 d_3 \over r^2}, ~~~~Z_2 \approx  {Q_2 \over r }, ~~~~Z_3 \approx  {Q_3
\over r }
\ee
and the leading terms in (\ref{k2}) are  
\be
k  \approx {d_2 Q_2 + d_3 Q_3 \over r^2}
\ee
Equation (\ref{ctc}) is only satisfied when 
\be
(d_2 Q_2 + d_3 Q_3)^2 \leq 4 d_2 d_3 Q_2 Q_3 ,
\ee
which implies 
\be
Q_3 d_3 = Q_2 d_2. 
\label{frozen}
\ee
Thus the background is free of closed timelike curves only when the tubes have the 
property (\ref{ratio}), which also comes out naturally from the 
Born-Infeld analysis. The subleading terms in (\ref{ctc}) also put a lower bound on 
the charge densities, similar to the ones in the two charge case. Equation 
(\ref{frozen}) also implies that the local charge densities $Q_2$ and $Q_3$ cannot 
vary independently along the tube. 

The Ricci scalar near the branes diverges like $r^{-2/3}$;  however this probably 
signifies the presence of brane sources, and is not a pathology. In our U-duality frame 
the two charge supertube (which is regular in the D1-D5 duality frame \cite{LuninIZ}) 
has a Ricci scalar which diverges like $r^{-4/3}$, again signifying the presence of 
brane sources\footnote{This is not surprising; many D-brane solutions which have 
a divergent Ricci scalar are related by T-duality to D3 brane solutions which are smooth.}.

\subsection{Three charges and one dipole charge }

In the Born Infeld analysis of \cite{bk} this case is pathological. Indeed, consider 
a D4 - F1 $\ra$ D6 supertube, and try to add some D0 charge (the third charge) on 
the tube without turning on a D2 dipole moment. The D0 branes feel no force from the D4 
branes and F1 strings, but are repelled by the D6 branes. Thus, a
second dipole charge (D2 brane charge in this case) is needed to adhere the D0 
branes to the D4-F1 supertube. 

In the absence of a second dipole moment, the leading terms in the harmonic functions are
\be
Z_i \approx {Q_i \over r}
\ee
while 
\be
k \approx {Q_1 d_1 \over r^2 }
\ee
Thus, for small enough $r$, $k^2$ will always be bigger than $Z_1 Z_2 Z_3$, and so 
the background always has closed  timelike curves. Therefore the tubes with 
three charges and only one dipole charge are not good solutions, as expected from the 
Born Infeld analysis.

We should also note that if one does not identify $z$ with an angular direction but 
keeps it infinitely extended, the closed timelike curves do not appear, and the metric 
is perfectly
well behaved. The curvature does diverge near the tube, but this might be again
the result of brane sources, and not a problem.


\section{The generic case -- supertubes and black supertubes}

In the general case, when the three charges and three dipole charges are nonvanishing, 
the asymptotic form of the metric near the tubes is:
\be
ds^2 \approx -{r^4 \over R_0^4} dt^2 + 2 {r \over R_0} dt dz +  c_z dz^2 + {R_0^2 \over r^2}
dr^2 + R_0^2 (d \theta^2 + \sin^2 \theta d \phi ^2) + ds^2_{T_6}   
\ee
where $R_0 \equiv 2 (d_1 d_2 d_3)^{1/3} $, the metric on the 6-torus is flat, and
\bea
R_0^4 c_z &=& - 16{{d_1}}^2{d_2}{d_3} - 16{d_1}{{d_2}}^2{d_3} 
- 16{d_1}{d_2}{{d_3}}^2  + \label{c}\\
&+& 2{d_1}{d_2}{Q_1}{Q_2}  + 2{d_1}{d_3}{Q_1}{Q_3} + 2{d_2}{d_3}{Q_2}{Q_3} 
- {{d_1}}^2{{Q_1}}^2 -  {{d_2}}^2{{Q_2}}^2 - {{d_3}}^2{{Q_3}}^2   \nonumber .
\eea 

The metric has a horizon at $r=0$, and a transverse 2-sphere of fixed radius $R_0$. 
The Ricci scalar at the horizon is constant, and proportional 
to the inverse square of the sphere radius. 
\be
R_{\rm scalar} \approx {1 \over 2 R_0^2} 
\ee

Thus, for large $R_0$ the geometry is smooth. 
The physics of the solution depends crucially on the sign of $c_z$. If 
$c_z$ is negative, the background has closed timelike curves, and is nonphysical. This
introduces a lower bound on the charge densities. As we will see in the 
next subsection, for $c_z =0 $ the geometry is the large radius limit of the metric 
of a regular three charge supertube. For $c_z > 0$, the horizon area is nonzero, and the 
geometry is that of a black string. This metric should be the large radius limit of 
the metric of a black supertube, which should reduce to a 5 dimensions BPS black ring.

\subsection{Regular three charge supertubes}

When $c_z =0$ the metric becomes
\be
ds^2 \approx -{r^4 \over R_0^4} dt^2 + 2 {r \over R_0} dt dz + {R_0^2 \over r^2}
dr^2 + R_0^2 d \Omega_2^2 + ds^2_{T_6}   
\ee

One can introduce new coordinates $u,v$ 
given by
\be
t = -u+v , ~~~ z = u+v , 
\ee
and a new radial coordinate $r = {\rho^2 \over 8 R_0}$. Ignoring 
subleading terms, the metric becomes 
\be
ds^2 \approx {\rho^2 \over 4 R_0^2} (-du^2+dv^2) + {4 R_0^2 \over \rho^2}
d\rho^2 +  R_0^2 d \Omega_2^2 + ds^2_{T_6}   
\ee
Thus, the metric in the near tube region is locally $AdS_3 \times S^2 \times T^6$.

Since the $AdS$ metric can be continued smoothly past $r=0$, and neither the Ricci 
scalar nor other invariants diverge there, the metrics of these three charge configurations are
completely smooth. This is an important test these solutions must pass if 
they are to represent the hair of the three-charge black hole, as conjectured in 
\cite{mathur,MathurSV}. 

The size of the $AdS_3 \times S^2$ region is of order the Planck scale 
for small $d_i$, and larger for tubes with larger M5 dipole charges. 
Our brief investigation shows that this metric can be continued 
behind $r=0$, has at most a zero area horizon, no closed timelike curves, and no
singularities. Thus, it has the right properties to describe a three charge supertube. There is
another way to see that this metric should indeed describe a supertube: 
When $Q_1=Q_2=Q_3 =Q$ and $d_1=d_2=d_3=d$, the 
condition $c_z =0 $ becomes $Q^2 = 16 d^2$. This condition is very natural if we think about 
the three charge supertube as being made of three simple tubes with two charges. The charge
densities of each tube are $Q/2,Q/2$. As we have seen in (\ref{2c}), a two charge tube 
has densities  $Q_i Q_j = 4 d_k^2 $, which implies  $Q^2 =16 d^2$, and hence $c_z=0$. 
Thus, it appears that the $c_z=0$ is a natural property of the supertube with three charges and 
three dipole charges.

This solution 
is reminiscent of the M3 brane of \cite{cvetic}, which is also made of three M5 branes preserving
4 supercharges \cite{im}, or of the M-theory $AdS$ solutions found in 
\cite{gaunt}. It would also be interesting to see if there exists a two dimensional 
conformal field theory dual to M-theory on this space.

\subsection{Black supertubes}

When $c_z$ is greater than zero, the area of the horizon at $r=0$ is larger than 
zero. Indeed, the horizon is extended in the $z,\theta,\phi$ directions, 
and $g_{zz}~ g_{\theta\theta}~ g_{\phi\phi}$ is finite at $r=0$. Thus, the metric 
for $c_z > 0$ describes the large radius limit of a BPS black tube (which should 
descend to a black ring in 5 dimensions).

Near the horizon the metric is
\be
ds^2 \approx -{r^4 \over R_0^4} dt^2 + 2 {r \over R_0} dt dz + c_z dz^2 + {R_0^2 \over r^2}
dr^2 + R_0^2 d \Omega_2^2 + ds^2_{T_6},   
\ee
which after completing the squares and ignoring subleading terms can be put in the form: 
\be
ds^2 \approx  - {r^2 \over R_0^2 c_z} d t^2 + c_z \left(d z + {r \over R_0 \sqrt{c_z}} 
d t \right)^2 
+ {R_0^2 \over r^2}
dr^2 + R_0^2 d \Omega_2^2 + ds^2_{T_6},   
\ee

Although it is not apparent, this metric is of the form $AdS_3 \times S^2 \times T^6$, 
with identifications
coming from the fact that $z$ was originally an angular coordinate \footnote{I thank 
Harvey Reall for pointing this out, and correcting a mistake in the 
previous version of this paper}. At first glance 
the horizon at $r=0$ is smooth, and the local $AdS$ coordinates can be extended behind the 
horizon in the usual way.

This black tube was conjectured to exist in \cite{bk}, based on the understanding of the 
black hole microstates put forth in \cite{mathur,MathurSV}. Indeed, if small angular momentum 
supertubes are to be the hair of the BMPV black hole, it is but natural to 
expect that very large tubes, whose angular momentum is larger than the angular momentum 
of the BMPV black hole, represent the hair of another BPS black object, which should be 
a black ring.

The fact that we have found near-tube solutions with a nonzero horizon area is a rather 
compelling piece of evidence that such a black ring exists. We should note that 
this three charge BPS black ring would be the first of its kind, since other known black 
rings \cite{EmparanWN,EmparanWY,h1} are nonextremal, or have pathologies \cite{ElvangMJ}.

\subsection{ Black Hole Physics}

For the purpose of discussing black hole physics it is convenient to 
look at tubes with three equal charges and three equal dipole charges 
($Q_1=Q_2=Q_3 =Q$ and $d_1=d_2=d_3=d$) to make the arguments 
more transparent.

As we have also argued, it is very natural to assume that the 
relation $c_z=0$ is a fundamental one for 
a generic three charge supertube, much like the relation $\tilde Q_1 \tilde Q_2 = 4 d^2$ 
is for a two charge supertube. In the equal charge case, these two relations are 
equivalent, and imply $Q = 4 d $. 
\begin{center}
\begin{figure}[h]
\centerline{\scalebox{.6}{\includegraphics{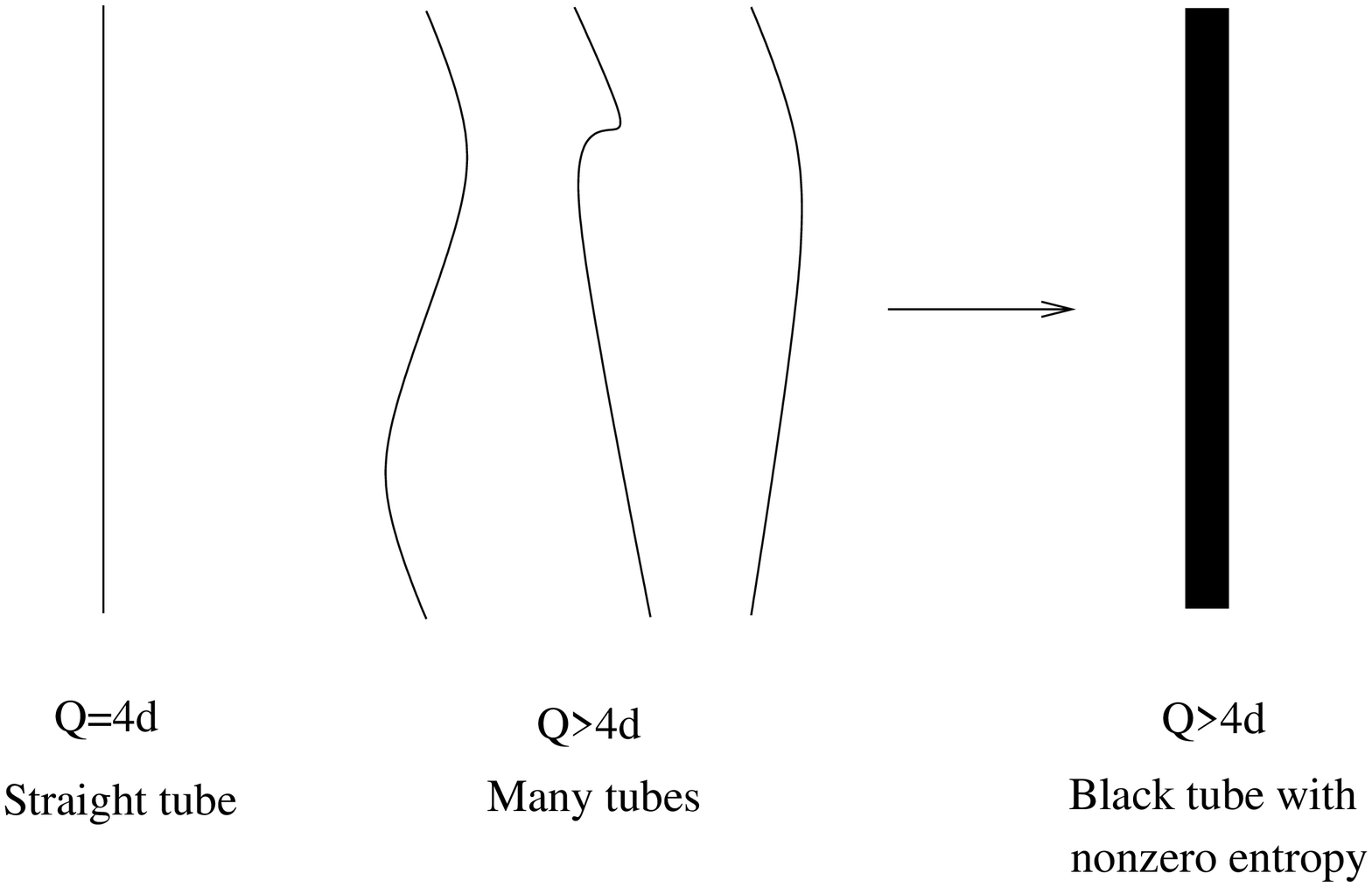}}}
\caption{As one increases $Q$, the multiplicity of supertubes and the horizon area of the black
tube increase.}
\end{figure}
\end{center}
\vspace{-1cm}

The picture which emerges from the above analysis is rather fascinating. If the local 
charge densities on a tube are such that $c_z=0$ (which in the equal charge
 case implies $Q=4d$) the only spherically symmetric solution is a 
straight supertube. If the charge densities are higher, the only spherically symmetric solution 
is a black tube. However, we know from the Born Infeld analysis that supertubes do not have to
be straight. Moreover, a wiggly  supertube has the charge to dipole charge ratio higher 
than a straight tube. Thus, if $Q > 4d$ there are also quite a few regular supertube 
configurations where the tubes are wiggly. As $Q$ gets larger, the length of the regular tube one
must fit in a fixed size box increases, 
and the tube can become more and more wiggly; hence, the number of configurations increases.

According to the interpretation of black hole horizons put forth in \cite{mathur,MathurSV}, 
the microstates of a black object are regular geometries, and the horizon 
is the place where these microstate geometries start differing from each 
other. Moreover, the horizon area counts the number of these microstates.

As argued in \cite{bk}, the microstates of the BPS black ring should be supertubes 
with angular momentum larger than that of the BMPV black hole.  Therefore, in the 
large radius limit we discuss, the 
microstates of the black tube are wiggly supertubes. For $Q = 4 d$, there is only one supertube 
configuration -- the straight one. This is consistent with the horizon area of the 
black tube being zero. As $Q$ becomes larger than $4d$, the tubes can become more and more 
wiggly, and their multiplicity increases. Moreover, their transverse size also increases. 
As the multiplicity and transverse size of the tubes increases, so does the horizon 
area of the black tube. 

The fact that a black tube forms exactly when the multiplicity of states becomes nonzero, 
and that the horizon area and the multiplicity grow together is a spectacular 
confirmation of the picture of black hole microstates of \cite{mathur,MathurSV}. This is 
further strengthened by the fact that the transverse displacement of the 
wiggly tubes (hence the size of the region where they start differing from each other) 
also grows\footnote{It would be interesting to try to find the geometries of the 
wiggly tubes and make this argument more precise.}.
Hence, the supertube geometries do exactly what the hair of the black tube should do. 

\section{Conclusions}

We have used some of the algebraic Killing spinor methods developed in \cite{nick} 
to find the 
large radius limit of the metric of three charge supertubes and BPS black rings. 
We have found
that the supertubes with three charges and two dipole charges obey a constraint on the 
local charge densities. This constraint is identical to the one found in \cite{bk} via the
Born-Infeld description of these supertubes.

We have also found that tubes with three charges and three dipole charges come in two types. 
If the charge densities and dipole charges satisfy a certain relation, the metric appears 
to be completely regular and there is no horizon -- this metric is that of a straight 
supertube. If the charge densities are larger, a horizon of nonzero area forms, and 
the metric is the near horizon metric of a black ring. 

We have also showed that the horizon area of the black ring and the multiplicity of the 
supertube geometries grow together, much as one would expect if the supertube geometries 
were the hair of this black ring, as predicted in \cite{bk} based on the picture of black 
hole microstates of \cite{mathur,MathurSV}.

There are quite a few directions to pursue. The first would be to use the 
algebraic Killing spinor methods used here to find the full metric of three charge supertubes 
and of three charge BPS black tubes. It is important to find these metrics, and to see if
 the most generic three charge supertube solutions are regular, and if they 
start differing from each other in a place where the horizon of the corresponding black hole would
be. It would also be interesting to 
fully explore the structure of the metric near $r=0$, and to find its analytical 
continuation behind the horizon.
 
Equations (\ref{har1}-\ref{har3}) allow one to look for a few rather interesting 
generalizations of our solutions. For example it might be possible to construct 
a solution with several three black holes joined by a flat supertube, 
or a black tube solution with variable charge densities.

Perhaps the most important application of our solutions is to use them as a testing 
ground for the conjecture of \cite{mathur,MathurSV}. Indeed, we have found a black 
tube whose horizon area is zero when there is only one supertube state, and increases 
as the number of supertubes increases. 
If one were able to explicitly count the three charge supertube configurations,  
(like Cabrera-Palmer and Marolf did in the two charge case \cite{bd}) and to 
explain the entropy of the black tube we would have a rather remarkable proof of 
the validity of this conjecture.

\bigskip
\noindent {\bf Acknowledgements:} 
\medskip

\noindent I'm deeply indebted to Nick Warner 
for help with the algebraic Killing spinor methods used to find this solution. 
I thank Per Kraus and 
Don Marolf for excellent suggestions, and comments on the manuscript. 
I have benefitted from useful discussions with Radu Roiban, Henriette Elvang and 
Belkis Cabrera-Palmer, whom I would also like to thank for correcting a sign 
error in equations (\ref{k0}-\ref{k}). 
This work is supported in part by the NSF grant 0099590.


\end{document}